\documentclass[twocolumn]{aastex631}
\usepackage{amsmath}
\usepackage{graphicx,subfigure}
\usepackage{hyperref}
\usepackage{parskip}
\begin{document}

\title{Calcium bright knots and the formation of chromospheric anemone jets on the Sun}

\author{Kunwar Alkendra Pratap Singh}
\affiliation{Plasma Astrophysics Research Laboratory, Department of Physics, Institute of Science, 
BHU, Varanasi-221005, India}
\affiliation{Astronomical Observatory, Graduate School of Science, Kyoto University, Yamashina, Kyoto 607-8471, Japan}

\author[0000-0001-6256-8628]{Keisuke Nishida}
\affiliation{Astronomical Observatory, Graduate School of Science, Kyoto University, Yamashina, Kyoto 607-8471, Japan}

\author[0000-0003-1206-7889]{Kazunari Shibata}
\affiliation{Astronomical Observatory, Graduate School of Science, Kyoto University, Yamashina, Kyoto 607-8471, Japan}
\affiliation{School of Science and Engineering, Doshisha University, Kyotanabe-city, Kyoto 610-0394, Japan}
\begin{abstract}
Space-based observations show that the solar atmosphere from the solar chromosphere to the solar corona is filled with small-scale jets and is linked with small-scale explosions. These jets may be produced by mechanisms similar to that of large-scale flares and such jets may be related to the heating of corona and chromosphere as well as the acceleration of solar wind. The chromospheric anemone jets on the Sun remain puzzling because their footpoints (or bright knots) have not been well resolved and the formation process of such enigmatic small-scale jets remains unclear. We propose a new model for chromospheric jets using the three-dimensional magnetohydrodynamic (MHD) simulations, which show that the continuous, upward rising of small-scale twisted magnetic flux ropes in a magnetized solar chromosphere drive small-scale magnetic reconnection and the launching of several small-scale jets during the evolution of the chromospheric anemone jets. Our new, self-consistent, three-dimensional computer modeling of small-scale, but ever-changing flux rope emergence in the magnetized solar atmosphere is fully consistent with observations and provides a universal mechanism for nanoflare and jet formation.
\end{abstract}

\section{Introduction} 
The small-scale jets, associated with small-scale explosions fill the solar atmosphere present between the solar chromosphere and to solar corona (Schmieder 2022).  There is increasing indirect evidence that these jets may be produced by a physical mechanism similar to that of large-scale flares, i.e., the release of magnetic energy stored in the solar atmosphere through magnetic reconnection (e.g. Yokoyama \& Shibata 1995, Shibata \& Magara 2011, Sterling 2015), and also that these jets may be related to the heating of corona and chromosphere as well as the acceleration of solar wind (Raouafi 2016, Sterling \& Moore 2020). Among small-scale jets, the so-called chromospheric anemone jets has a inverse Y-shape structure in active regions on the Sun remain puzzling, because they are at such small scales that their footpoints (bright knots) have not been well resolved (Shibata et al. 2007, Nishizuka et al. 2011, Singh et al. 2012a,b).

In addition to the chromospheric anemone jets, the solar chromosphere harbors a variety of jet-like phenomena such as spicules (Zhang et al. 2012), penumbral microjets (Katsukawa et al. 2007), light bridge jets (Asai, Ishii \& Kurokawa 2001; Shimizu et al. 2009), IRIS bombs (Tian et al. 2016). Chromospheric anemone jets usually occur in solar active regions and the name \textit{anemone} is derived from the morphology of coronal X-ray jets reported earlier (Shibata et al. 1994). The apparent velocity of the chromospheric anemone jets ranges between 5~ km s$^{-1}$ to 20 ~km s$^{-1}$, the lifetime of these jets ranges between 100 s to 500 s, and the apparent velocity of chromospheric anemone jets deduced from the SOT/Hinode observations lies somewhat close to the local Alfv\'en speed in the lower solar chromosphere (Nishizuka et al. 2011). Sometimes the jets are found to be invisible in the Calcium-$K_3$ (or Ca-$K_3$) component (Morita et al. 2010). Since the Ca-$K_3$ component is emitted from the top of the solar chromosphere, some of the jets that were reported earlier correspond to the lower chromospheric phenomena (Morita et al. 2010). However, giant chromospheric anemone jets are also reported in observations and such jets appear to be associated with the reconnection in the upper solar chromosphere or solar transition region (Nishizuka et al. 2008). 

The chromospheric anemone jets are collimated plasma ejections, and, below the jet's spire, there is a presence of a bright loop or a bright knot. In SOT/Hinode observations, the bright knots are visible in Ca {$\scriptsize II H$} broadband filtergram images. There are blob-like, multiple plasma ejections that occur in the bright knots and develop gradually into a narrow, collimated jet (Singh et al. 2012a), usually appearing as a single thread at least within an arcsec (Nishizuka et al. 2011, Singh et al. 2011, Singh et al. 2012a). It is believed that the chromospheric anemone jets are formed as a result of the magnetic reconnection between an ambient magnetic field and an emerging magnetic bipole (Shibata et al. 2007), and, if the magnetic reconnection occurs below the middle solar chromosphere or solar photosphere, it indeed leads to the formation of jets that appear to correspond to chromospheric anemone jets (Takasao, Isobe \& Shibata 2013). Jets on small- scales in the solar chromosphere are very much intermittent in nature (Singh et al. 2012a, Tian et al. 2014), however, there is no convincing, self-consistent picture that establishes a link between the emergence of bright knots, the launching of small-scale jets from the footpoint region or bright knots during the evolution of chromospheric anemone jets. The 2D magnetohydrodynamic (MHD) models alone cannot produce many jets from the entire body of magnetic bipole, because, in 2D MHD, the jet ejection is \textit{localized}. In case of 2D MHD, the magnetic reconnection occurs primarily on one side of the footpoint and does not erupt during jet activity. Therefore, we need a new 3D model in order to explain enigmatic features linked with the chromospheric jets. 

\section{Observations}
In the present work, we report a high-resolution Calcium {$\scriptsize II H$} filtergram (FG) image of a chromospheric anemone jet that was observed by SOT/Hinode (Tsuneta et al. 2008, Suematsu et al. 2008, Ichimoto et al. 2008) aboard Hinode (Kosugi et al. 2007), on 2006 November 20 in NOAA AR10923, in the northern hemisphere, with a time cadence of about 8~s, and having a $\lambda$-shaped footpoint from SOT/Hinode. 

Figure 1a shows fine-scale, enigmatic structures of one of the chromospheric anemone jet (Singh et al. 2012b) and the formation mechanism of such features in the solar chromosphere is still unclear. 

In a well-developed Calcium bright knot that is larger than usual, one can notice fine structures along with a well developed jet spire (main jet). Figure 1b shows running difference image of the small region indicated by the box in Figure 1a at 12:40:22 UT -12:40:14 UT to show the small-scale jets emanating from the bright knots. 

Figure 1c shows the snapshot of the small-scale jets and a $\lambda$-shape region above the bright knots at 12:43:17 UT. Several small-scale jets appear from the Calcium bright knots. The small-scale jets develops further in to the solar atmosphere and move systematically from one end of the bright knot. Later on, these small-scale jets merge into the main jet (see e.g. Figures 1b, 1c, 1d along with online movie1). The main jet is clearly visible in the running difference images, derived from Calcium filtergram images of SOT/Hinode (see e.g. Figure 1b, 1d and online movie1). The unsharp masking is applied to Fig. 1a and 1c. The The faint jets as well as multi-thread structures are clearly visible in the running difference image (Figure 1d). \href{https://www.kwasan.kyoto-u.ac.jp/\~nishida/singh2024/movie1.mp4}{The online movie1} provides the evolution of the jets between 12:00~UT to 13:59~UT. The left panel in the online movie1 is obtained by performing the unsharp masking, the middle panel is obtained by applying the derivative filter, and the right panel is obtained by applying the running difference method.

The launching positions of the main jet and the small-scale jets are not the same. The small-scale jets are launched from various locations of the bright knot that is an extended structure. The jet activity near the bright knots become pronounced and several small-scale jets are launched from the Calcium bright knots, after the main jet is developed. The systematic motion of small-scale jets are seen more clearly when the pieces of bright knots merge into each other, and form a rather extended structure, at some stage. 

The top end of the jet (or the jet spire) in the Calcium image provides the information of the apparent velocity of the jet. The jet velocity of the main jet, at a later stage is $\sim$ 20 km/s. 

Figure 1e shows the height-time plots of the brightness of small scale jets along the slit (long rectangle in Figure 1a). In the height-time plots, we are interested mainly in the small-scale jets located near the footpoint region or the bright knots. The slit is made wider to accommodate the chosen jets inside the slit. The analysis of the height-time diagram of the jets near a selected portion of the Calcium bright knot shows several small-scale jets emanating from the footpoint, along with a presence of fine-scale features in the footpoint. The very fast velocities indicated by the dashed white line, in Figure 1e, is not real but the apparent motions due to the transversal motion of the jets inside the slit. Nevertheless, the small-scale jets emanating near the Calcium bright knots have higher velocities ($\sim$ 40 km/s).
\section{Three Dimensional MHD Simulation of Chromospheric Anemone Jets}
%%hypothesis
To explain enigmatic chromospheric anemone jet shown in Figure 1, Singh et al. (2012b) proposed the hypothetical model in which many small scale jets are generated by the successive magnetic reconnection associated with the emergence of twisted flux tube (as shown in Fig. 4c of Singh et al. 2012b).

Here we try to develop above Singh-model using 3D MHD simulations of reconnection associated with emergence of twisted flux tube in the solar active regions.
\subsection{Details of Numerical experiment}
Our model is a direct extension of the 2D MHD model of the coronal jets and surges developed by Yokoyama \& Shibata (1995, 1996). In our self-consistent, 3D MHD simulations, the initial gas layer contains the sub-photosphere (convection zone or CZ), solar photosphere, chromosphere, transition region, solar corona. All the gas layers are in the hydrostatic equilibrium under uniform gravity. The CZ gas layer is convectively unstable. The initial temperature at the solar photosphere is set to 6000 K, whereas the initial temperature in the solar corona is set to 6 $\times$ 10$^5$ K. The temperature distribution is akin to the solar atmosphere. Figure 2 shows the initial distribution of magnetic field and size of the computational box. Note that $z=0$ is assumed to be the base of the solar photosphere. 

%% initial magnetic field
%%value of magnetic field
The background magnetic field is uniform and parallel in the vertical direction (Figure 2). A weakly twisted magnetic flux rope is inserted below the solar photosphere. See Gold \& Hoyle (1960) for the initial magnetic flux rope configuration utilized in our simulations. The axis of the twisted flux tube is in the x-direction, the positive z-direction is vertically upwards and the y-direction is perpendicular to x- and z-directions but lies on the xy-plane. 

The initial flux rope configuration is kink unstable. The initial values for the coronal magnetic field are taken to be 370 G and the magnetic field of the flux rope at the centre and at $z= -1100$~km  is taken to be 5600 G. The twist in the magnetic flux rope is in the anti-clockwise (right-handed) direction. 

A twisted magnetic field configuration has to be utilized for the emerging magnetic fields in the simulations due to the fact that the S-shaped, sigmoid structures are often associated with large-scale flux emergence.  Solar jets are often associated with spinning motions (Canfield et al. 1996), so there always exists a possibility of twisted flux tubes at small scales. Sometimes the twisted magnetic fields can be formed as a result of local sheared configuration (e.g. Takasao et al. 2015). The possibility of wave excitation due to the small-scale, reconnection-generated jet formation in relation to the chromospheric heating cannot be ignored (Kotani \& Shibata 2020).
%% solving MHD equations
%% and CIP MOCCT

We have examined the emergence of twisted flux tube in solar atmosphere and interaction (reconnection) with ambient field, by solving the 3D resistive MHD equations. We assume adiabatic energy equation and consider resistivity in the magnetic induction equation. We assumed anomalous resistivity depending upon the current density for causing fast reconnection (e.g., Yokoyama and Shibata 1996, Nishida et al. 2013). The heating term due to the resistivity is neglected in the energy equation. The ratio of specific heat is 5/3.

%% Size of comutational box, box size and grid points, boundary conditions
The boundary condition is free at top and side boundaries. The bottom boundary is assumed to be slipping wall boundary (so that magnetic field and plasma cannot pass through there but move freely along the boundary). 

%%Initial perturbation
We did not assume convective flow initially in the simulations. Rather, a slight upward velocity field is added to the centre of the flux rope. The initial velocity distribution, between -550~km $< x <$ 550~km, is: $V_z$ = $v_0$ cos[($\pi$/10)($x/H$)]~, where $v_0$ = 7.1 $\times$ 10$^{-3}$~km s$^{-1}$, and $H$=110~km. Here $v_0$ can be considered as the minimum enough value to trigger the emerging flux in the absence of solar convection zone.
The numerical computation is carried out using the CIP-MOCCT scheme (Kudoh, Matsumoto \& Shibata 1999).
\subsection{Numerical Results}
Figure 3a shows the density distribution and provides a simulated image of jets from our 3D MHD simulations. It is quite evident that the morphology of the entire jet structure appears differently when the structure is observed from two different viewing directions (Figure 3a). Figure 3b provides the temperature distribution of the jet.  The 3D computer simulation provides the clear information of the region present below the jet. \href{https://www.kwasan.kyoto-u.ac.jp/\~nishida/singh2024/movie2.mp4}{An animation (movie2) of Figures 3a and 3b is available in digital version.} The movie2 provides the jet evolution obtained using density distribution (upper panel) from the simulation data. The region below the jet shows a clear anemone (or inverse Y) shape when viewed along the axis of the twisted magnetic flux rope. The online movie2 also provides the jet evolution obtained using temperature distribution (lower panel) from the simulation data. The loop or the extended structure (footpoint) present below the jet has higher density compared to the jet. The temperature of the footpoint region below the jet has a cooler and a hotter component, although the majority of the footpoint structure present below the jet has a cooler component. Such a footpoint region having a higher density and multi-thermal temperature structure, compared to the overlying jet structure, could be visualized as a bright loop or bright knot that are reported quite often in the SOT/Hinode observations. 

Figures 3c and 3d provide the absolute velocity of the jets in the yz-plane for $t=1.94 \times 10^{3}$~s. Figure 3c corresponds to $x=0.028$~Mm (i.e. near the centre of the twisted flux tube) and Figure 3d corresponds to $x=8.8$~Mm (i.e. away from the centre). The velocity distribution of the jet near the centre of the flux tube i.e. $x=0.028$~Mm shows that the jet velocity reaches a value of about 100 km s$^{-1}$. The velocity distribution of the jet in the outer region i.e. $x=8.8$~Mm is in the range of 7 km s$^{-1}$ to 15 km s$^{-1}$. 

Our 3D MHD simulations show that the continuous, upward-rising motion of the magnetic flux rope creates multiple reconnection sites and results in the jet formation at various locations and heights in the solar atmosphere. Many small-scale jets are launched from the body of the flux rope, develop gradually towards the main jet and merge with the main jet. The evolution of the main jet and small-scale jets in the computer simulations are very much similar to the observations of the jets. A weakly twisted magnetic flux rope provides a guide magnetic field, and the presence of this axial component, during the upward rising motion of the flux rope, is vital for the launching of many small-scale jets and their systematic motions from one end of the bright knot to the other. 

The chromospheric anemone jets are embedded in the magnetic field of the continuously emerging magnetic flux rope. Previous SOT/Hinode observations show that after the launching of jet activity, the bright knots or anemone-shaped footpoint start to shrink with time (Singh et al. 2012a), and it is quite likely that a major portion of the Calcium bright knots, present below the jets, correspond to emerging flux. 

In the present observation, the bright knots do not appear at once. The different parts of the footpoint (or bright knot) appear in stages, form an extended structure later on, and eventually gives a $\lambda$ -shape to the jet. Figure 4a provides the height-time plot of the high-speed, small-scale jets formed near the bright knots in observations. The height-time plots corresponding to Figure 4a, obtained from 3D MHD simulation, are provided in Figure 4b and 4c. The bright parts near the footpoint appear with a speed of about 6.1~km s$^{-1}$ or slightly larger. The emerging flux region in the simulated height-time plot shows motions of about 8.6~km s$^{-1}$ which is comparable to the upward speed ($\sim$ 6.1 km s$^{-1}$) of the rising motion of the bright knot at the footpoint of the jets. The 3D MHD simulations clearly show vertical jet velocities as high as 80 km/s.

Fig. 5a shows the magnetic field structure at $t=1.94 \times 10^{3}~s$. The magnetic field lines at $t=1.94 \times 10^{3}~s$ show the reconnected magnetic field, along with the magnetic fields of the flux rope and plasmoid formed by magnetic reconnection. It is clear that the reconnected field lines are twisted. Fig. 5b shows the magnetic field and the velocity field at $x=0.028$~Mm and $t=1.94 \times 10^{3}~s$, and during this time, the x-point is located around $z=1.5$~Mm. The time evolution of the magnetic field and velocity show the presence of high-speed jets (e.g. around~40~km s$^{-1}$ to 60 km s$^{-1}$) in the computer simulations. It has been found that the height of the x-point influences the speed of the jet strongly, and the high-speed jets are formed due to the magnetic reconnection occurring at around $z=1.5$~Mm in the solar atmosphere. The schematic diagram i.e. Figure 5c to Figure 5j shows the formation and ejection of jets from the body of flux rope, as viewed from two different viewing directions. 
\section{Discussion and Conclusion}
Understanding the jet velocity, especially the faster velocity components ($\sim$ 40 km s$^{-1}$ - 100 km s$^{-1}$ or more) present in the jets, is fundamental. The jet velocity ($\sim$10 km s$^{-1}$) corresponding to the outer region, in our simulations, is comparable to the velocity of chromospheric anemone jets (Shibata et al. 2007, Nishizuka et al. 2011), though, in observations, velocities faster than 40 km s$^{-1}$ are also reported in some cases (Nishizuka et al. 2011). It is interesting to note that velocities around 100 km s$^{-1}$ are found near the centre of the flux tube in our simulations, much faster than those of typical chromospheric anemone jets. On the other hand, in one observation (Nishizuka et al. 2008) where the size of the emerging bipole was relatively large, the jet was accelerated to around 100 km s$^{-1}$. Overall, a careful estimate of the small-scale jet velocities in our particular observations, near the Calcium bright knots, gives a higher value compared to the usual values reported earlier during the statistical studies of chromospheric anemone jets (e.g. Nishizuka et al 2011). The apparent velocity of small-scale jets through eye-detection is $\sim$ 40 km s$^{-1}$ - 100 km s$^{-1}$, but higher velocity components are also possible.  

The Calcium bright knots before the appearance of jetting are studied in observations. The high-resolution Calcium filtergram observations from SOT/Hinode have revealed that there exists a significant time lag between the bright point or Calcium loop/knot and the appearance of the main jet, see e.g. Table 1 of Singh et al. (2012a), where SOT/Hinode observations show that the time- lag between the appearance of the bright loop and appearance of the main jet ranges between 1 min- 10 min, in fact, the time-lag between 1.5~min - 3~min is more common. In our present observations where the Calcium bright knot appears in parts that merge gradually, the time-lag is not fixed with typical time-lag between ~3~min - 4~ min. It is interesting to see that in our 3D MHD simulations, the typical time- lag between the appearance of the loop and the main jet is $\sim 3.2$ min. The time-lag noticed in our computer modelling should also be applicable to emerging flux and jets.

Multiple jets could be launched from a location if there is a recurrent emergence of bipoles in a small region of the solar chromosphere. In the present work, recurrent jets are produced in our 3D MHD simulations, and we provide a different scenario where there would be a presence of a single, dynamically evolving bright knots (or bipole), and many small-scale jets are launched as a result of multiple magnetic reconnection occurring in the entire body of the bright knots, and such a scenario fits very well with the observations of the chromospheric anemone jets. There are jets associated with light bridges that are observed in the umbral region of the sunspots (Shimizu et al. 2009, Shimizu 2011, Asai, Ishii \& Kurokawa 2001, Schmieder 2022). The morphology of such jets is similar to the successive ejections or small scale jets associated with chromospheric anemone jets, but the latter occur much more frequently than the light bridge jets.

Interestingly, in our self-consistent simulations, our new model can describe jets not only occur near the mixed polarity region (at the level of solar photosphere) but elsewhere in the body of the flux rope, exhibiting a thread-like pattern, and the footpoint continues to evolve in the upward direction and horizontal expansion takes place during the launching of the small-scale jets. The velocity distribution of the jet on the yz-plane at different x-positions in the simulation domain provides different values of the jet velocity. While developing interpretations of solar jets that are based only on the examination of the observed morphological features, one should be cautious because the jet formation requires a proper understanding of the magnetic reconnection (Isobe et al. 2005; Isobe, Proctor \& Weiss 2008) along with the driving mechanisms that can prevail at various heights in the solar atmosphere (Isobe et al. 2005; Shibata et al. 2007; Tian et al. 2014). Observations show that the nature of magnetic reconnection is quite bursty and time-dependent in solar chromosphere (Singh et al. 2012a). The height of the magnetic reconnection and interaction of slow mode shocks with the solar transition region is important (Takasao, Isobe \& Shibata 2013) for chromospheric jets and such effects should be studied in 3D MHD simulations.

The initial velocity trigger (by a small amount) is near the centre of the flux rope, so the middle part of flux rope will start rising and appear earlier in the higher atmosphere. The magnetic reconnection occurs also in the solar chromosphere in our 3D MHD model. As time goes on, the magnetic reconnection near the centre of the flux rope would take place at higher heights in the solar atmosphere, compared to regions of the flux rope that are located near the outer regions of the flux rope, and the main jet appears. In this situation, the small-scale jets would move towards the main jet. If location of the initial velocity perturbation is near the outer region of the flux rope (initial, twisted magnetic field), the outer part will emerge first at the higher heights, and one would then expect higher velocities of small-scale jets in the outer region, because the magnetic reconnection will occur at higher heights in the solar atmosphere. Even though the location of the initial velocity perturbation is important, the underlying physics would not be affected much.   

The faster velocity jets ($\sim 100$ km s$^{-1}$) could be due to the fact that the magnetic reconnection takes place in higher layers where the local Alfv\'en speed is high. The magnetic field and density both decrease with height. The ratio of magnetic fields in the upper and middle chromosphere is $\frac{B_{\rm upper}}{B_{\rm middle}} \sim \frac{1}{5}$ but the ratio of the number densities would be $\frac{n_{\rm upper}}{n_{\rm middle}} \sim 10^{-3}$. So, if we take the ratio of the local Alfv\'en speeds at two different heights i.e. upper chromosphere and middle chromosphere, the local Alfv\'en speed in the upper chromosphere should be 10 times higher than the local Alfv\'en speed in the middle chromosphere. Therefore, one would expect jets formed in the upper chromosphere to have higher velocities than jets formed below the middle chromosphere. 
 
We now discuss the effect of our assumption, especially adiabatic energy equations in 3D MHD simulations. The main reason of this simplified assumption is the difficulty of assuming realistic energy equations for chromospheric plasma, since radiation cooling is controlled by non-LTE radiative transfer. On the other hand, the gravitational stratification plays an important role in chromospheric jet formation and subsequent acceleration (e.g. Takasao, Isobe \& Shibata 2013). Even though we do not include radiation, we consider that the gravitational stratification provides a much stronger effect on the jet dynamics in the solar chromosphere. This is because the characteristic velocity of the jet is determined by the local Alfv\'en speed and the density determines the local Alfv\'en speed. The temperature can indirectly affect density distribution, e.g., if pressure balance is satisfied, and if the temperature increases twice, then density decreases to half, leading to an increase in Alfv\'en speed by $\sqrt{2}$. However, even if temperature changes, if density does not change, the Alfv\'en speed remains the same, so there is no influence on jet dynamics. The resolution ($\sim$ 55~km) is sufficient enough to resolve the small-scale jets. Our model is the first step of 3D MHD modeling of enigmatic chromospheric jets, and we believe, we can capture essential physics of dynamics associated with magnetic reconnection and emergence of twisted flux tube even under the assumption of adiabatic energy equation. The inclusion of realistic energy equation will be important future task.

The coronal physical quantities such as temperature, density, and gas pressure are not realistic values in our initial conditions, as in the 2D model of Yokoyama \& Shibata (1995, 1996). The temperature of solar corona is lower than typical coronal temperature in active region (i.e. 1~ $\times$ 10$^6$~K to 2~$\times$ 10$^6$~K ). The height of the solar transition region is much less than the realistic value. The density of coronal plasma is much greater than the realistic value (see e.g. Yokoyama \& Shibata 1996). Also see two-dimensional MHD simulations of Nishizuka et al. (2008) where a realistic coronal density and temperature is adopted compared to Yokoyama \& Shibata (1996), but it bring out similar physics. These demerits are not essential for the modeling of 3D jets. The presence of radiation in the energy equation would lead to a more realistic temperature distribution, though dynamics will remain unaffected, driven as it is by the Lorentz force. In the future, more realistic 3D modeling of jets will be made using large computational resources.

In conclusion, the present work provides insight through our new model of chromospheric anemone jets. For the first time, it has become very much evident, that the continuous, upward rising of the twisted magnetic flux rope in the magnetized solar atmosphere, is the key behind the formation of the chromospheric anemone jets, and the new MHD model has the capacity to resolve diverse, intricate issues related with the chromospheric anemone jets and its footpoint evolution. Under such circumstances, the jet activity in the solar chromosphere has to be understood from the viewpoint of the driving mechanisms and the underlying jet dynamics common to large-scale flares and mass ejections.

\section*{Acknowledgment}
Hinode is a Japanese mission developed and launched by ISAS/JAXA, with NAOJ as a domestic partner and NASA and STFC (UK) as international partners. It is operated by these agencies in cooperation with ESA and NSC (Norway). This research is supported by JSPS KAKENHI grant number 21H01131 (K. Shibata). KAPS gratefully acknowledges the UGC Faculty Recharge Program of the Ministry of Human Resource Development (MHRD), Govt. of India and University Grants Commission (UGC), New Delhi, the incentive grant of the Institute of Eminence (IoE) Program, BHU, and the Visiting Associateship Program of IUCAA, Pune. The numerical computations were carried out on SX9 at the Center for Computational Astrophysics (CfCA) of the National Astronomical Observatory of Japan, and the KDK system at the Research Institute for Sustainable Humanosphere (RISH) at Kyoto University. The authors are grateful to M. Shimizu, K. Uehara, and all the staff, members of the Astronomical Observatory of the Graduate School of Science, Kyoto University, Japan. We thank Prof. Marco Velli, Department of Earth, Planetary and Space Sciences, University of California, Los Angeles (UCLA) for constructive comments.
%%

%%
%%%%
%%%%%%%%%%
\begin{figure}
\label{j1}
\centering
\includegraphics[width=0.8\textwidth]{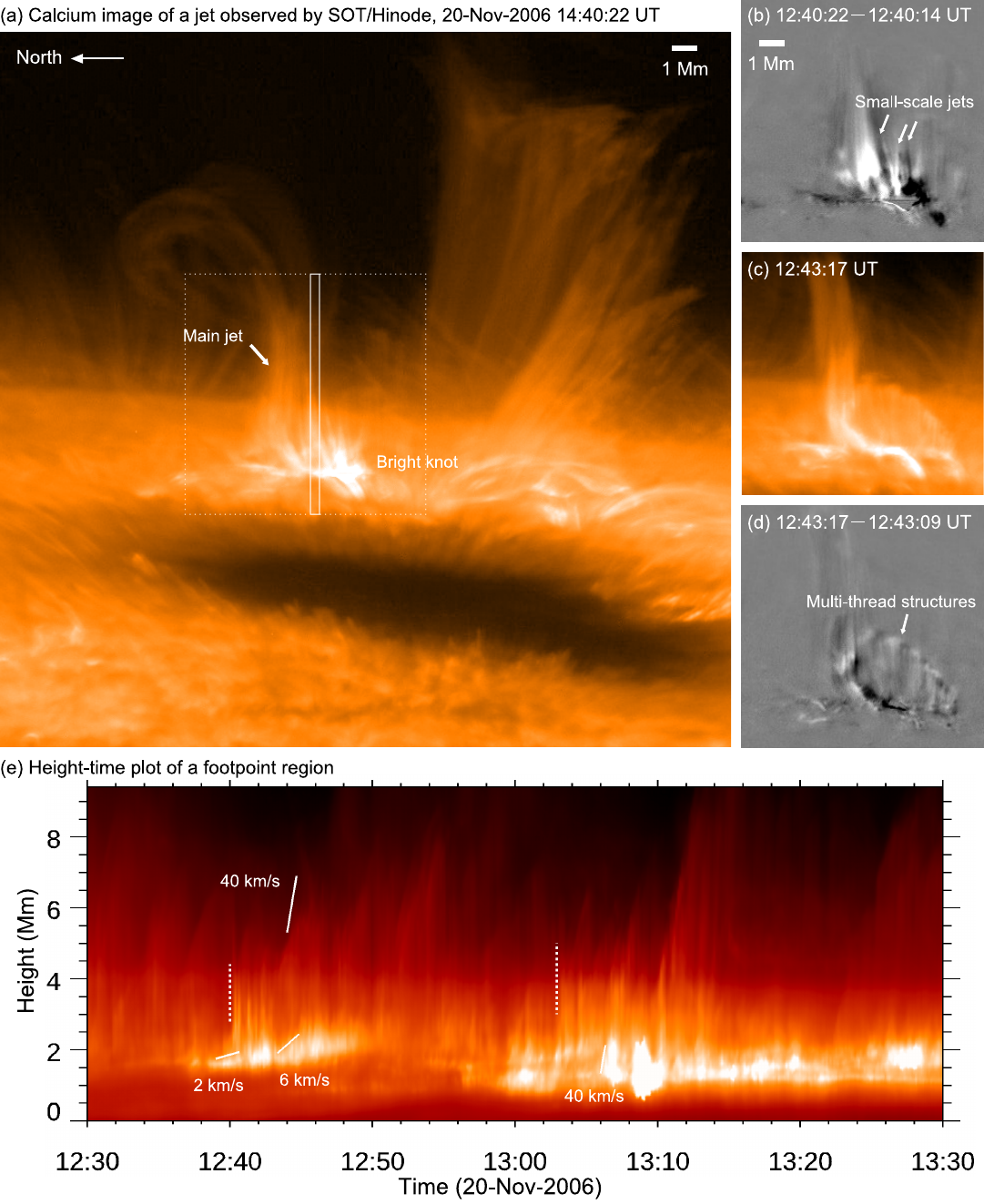}
\caption{ The small-scale jets from the bright knot in Ca {$\scriptsize II H$} broadband filtergram data obtained from high-resolution data from SOT/Hinode. The rectangular box with a solid line in the image shows the slit position. The rectangular slit is 10 pixels wide and it is utilized in the making of a height-time plot later on. (a) The main jet along with a bright knot in enigmatic solar chromosphere during high-resolution observations, as seen from SOT aboard Hinode satellite.  (b) Running difference image to show the small-scale jets emanating from the bright knots. The running difference images are derived from the observations of Calcium filtergram images taken from SOT/Hinode to show the small-scale jets jutting from Calcium bright knot, also see Singh et al. (2012b)  (c) Ca {$\scriptsize II H$} image to show the small-scale jets and $\lambda$-shape at 12:43:17 UT, (d) Running difference image to show that, along with a main jet, there exist small-scale jets along with multi-thread structures. The multi-thread structures at a later stage are more clear. The field of view of Figs. 1b to 1d are shown in Fig. 1a with a dotted rectangular box. Scales are provided for the sake of comparisons, (e) The height-time plot, derived from Calcium filtergram images (in observations), of a region below the observed jet that is present in the calcium bright knot (or footpoint). The rectangular slit position is shown in Fig. 1a. Typical velocities are shown with white ticks in the height-time plot. The very fast velocities indicated by dashed white lines, on Fig. 1e, are not real but apparent motion due to transversal motion of jets on the slit. Also, see \href{https://www.kwasan.kyoto-u.ac.jp/\~nishida/singh2024/movie1.mp4}{online movie1} where the Field-of-view (FOV) of the jets, obtained from the Ca {$\scriptsize II H$} filtergram data, is shown in Fig.1a with a dotted rectangular box, and jets arise between 12:00~UT to 13:59~UT.}
\end{figure}

\begin{figure}
\label{j5}
\centering
\includegraphics[width=0.90\textwidth]{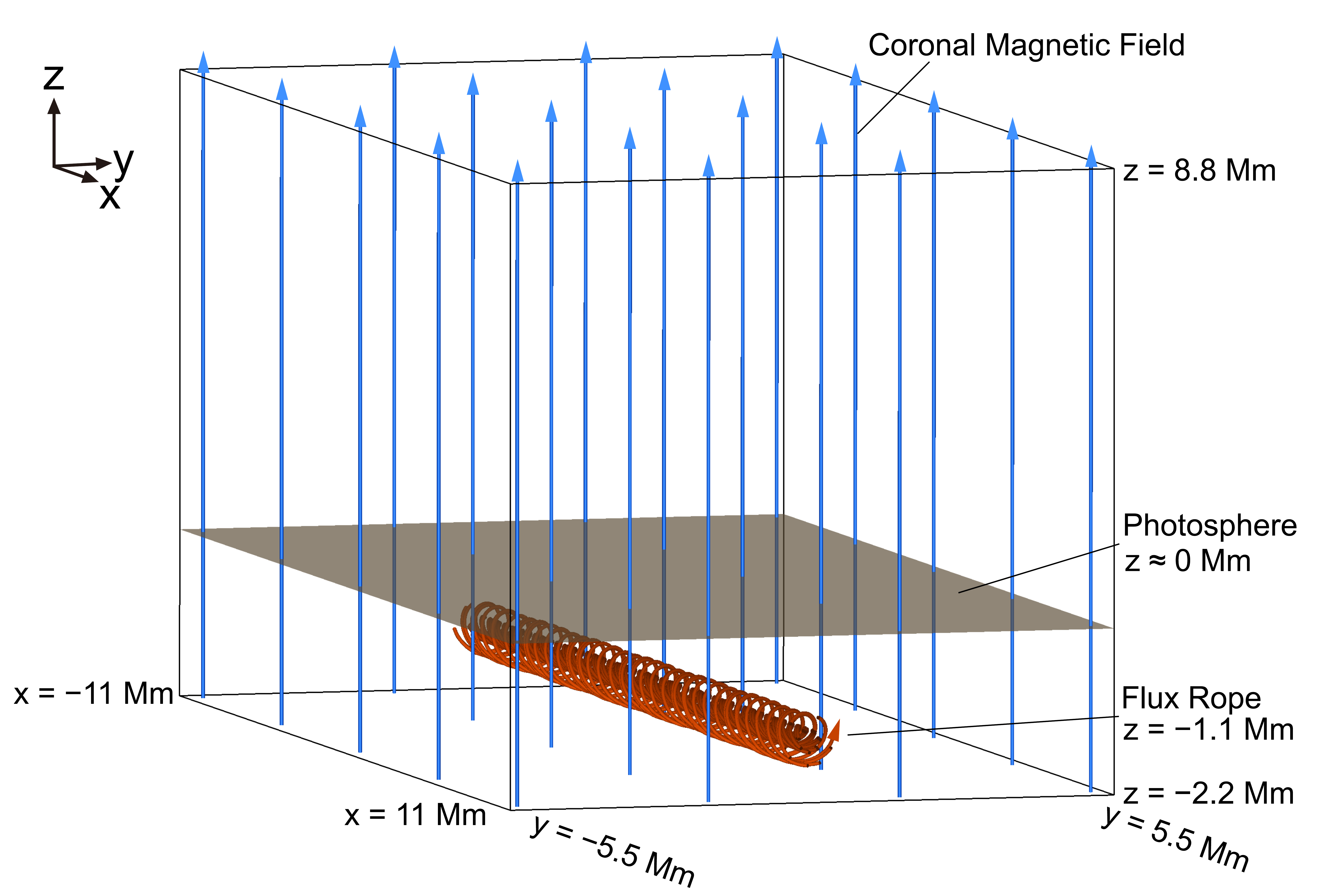}
\caption{The initial magnetic flux rope configuration along with the uniform, ambient magnetic field utilized in 3D MHD simulation. As time goes on, the initial magnetic flux rope configuration evolves dynamically during the upward rising and at some point in time, the jets are formed.}
\end{figure}

\begin{figure}
\label{j2}
\centering
\includegraphics[width=0.9\textwidth]{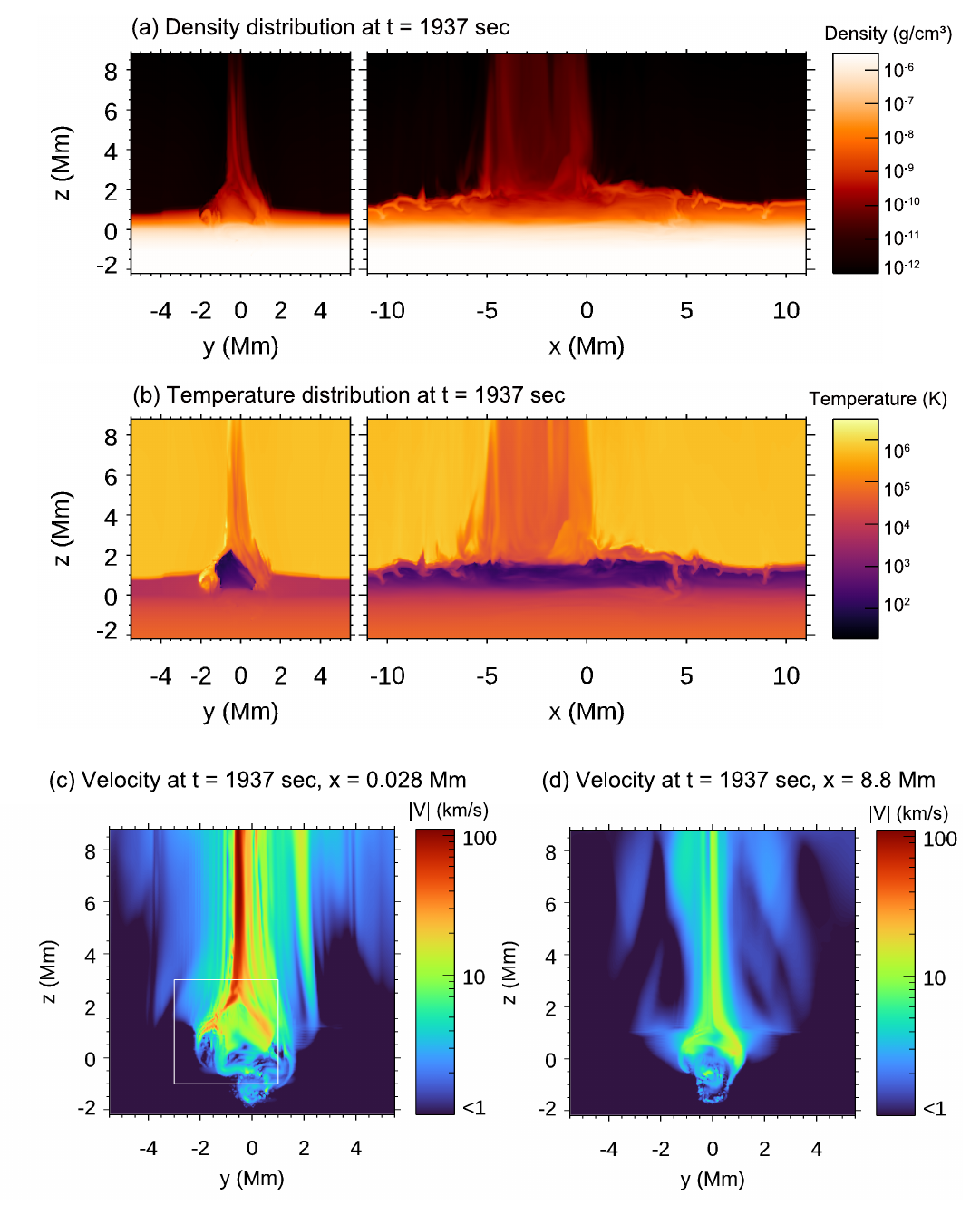}
\caption{The small jets along with the first jet (or main jet) are also noticed in 3D MHD simulations. (a) The density distribution from 3D MHD simulation shows several such jets launched from the footpoint. The density distribution at $t=1.94 \times 10^3 $~s is shown for two different viewing directions, (left: along the axis of the twisted magnetic flux tube, yz-plane, x = 0.028 Mm,(right: perpendicular to the axis of the twisted magnetic flux tube, xz-plane, y = 0.028 Mm (b) The temperature distribution at $t=1.94 \times 10^3$~s, $y = 0.028$~Mm for viewing direction perpendicular to the jet axis is shown.  (c and d) The absolute velocity distribution of the jets on yz-plane. The figures show two cases for $t=1.94 \times 10^3$~ s, panel (c): $x=0.028$~Mm, panel (d): $x=8.8$~Mm. The $x=0.028$~Mm inside the flux rope is the position near the axis of the flux rope and the jets have higher velocity components. In the outer region i.e. $x=8.8$~Mm the jets have relatively smaller velocity components. The white box in Fig. 3(c) corresponds to the region shown later on in Fig. 5(b). \href{https://www.kwasan.kyoto-u.ac.jp/\~nishida/singh2024/movie2.mp4}{An animation (movie2) of Figs. 3a and 3b is available in the digital version.} The upper panel in movie2 provides the time evolution of jets using density distribution obtained from 3D MHD simulation data. The lower panel in movie2 provides the time evolution of jets using temperature distribution obtained from 3D MHD simulation data.}
\end{figure}

\begin{figure}
\label{j3}
\centering
\includegraphics[width=0.60\textwidth]{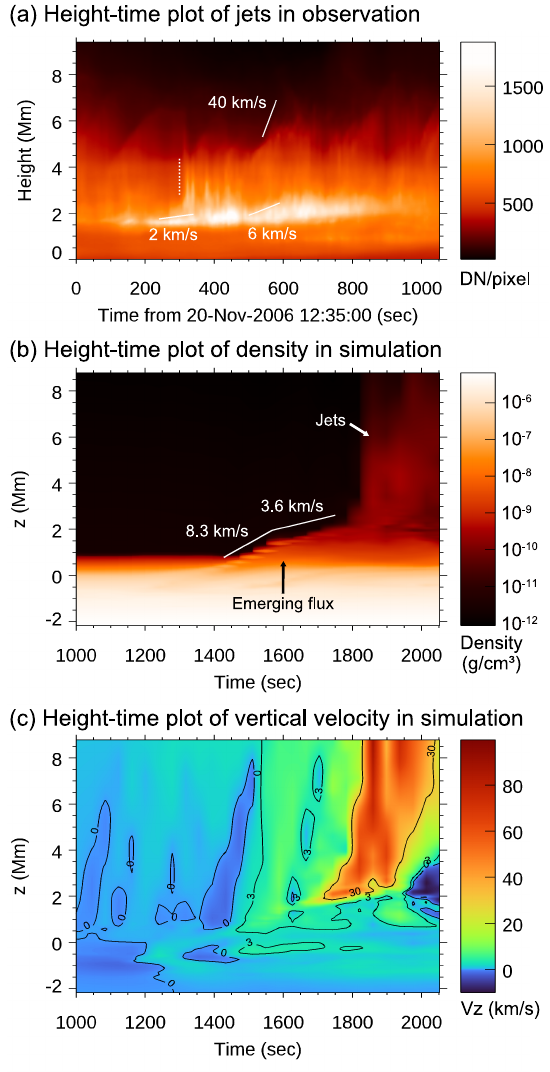}
\caption{ (a) The height-time plot, derived from Calcium {$\scriptsize II H$} filtergram images a selected region near the Calcium bright knot. The rectangular box is the same as the one used for Fig. 1e. Velocities are shown with white ticks in the height-time plot. The observed height-time plot can be utilized for comparisons with MHD simulations. The very fast velocities indicated by dashed white lines are not real but apparent motion due to transversal motion of jets on the slit, (b) a simulated height-time plot, derived from 3D MHD modeling, of the jet using the density distribution, (c) a simulated height-time plot of the jet using the velocity distribution. The contours in the velocity distribution shows the velocities in units of km s$^{-1}$. Both (b) and (c) correspond to $x=0.58$~Mm, $y=-0.47$~Mm. }
\end{figure}
%%
%%

%%\newpage
\begin{figure}
\label{j4}
\centering
\includegraphics[width=0.90\textwidth]{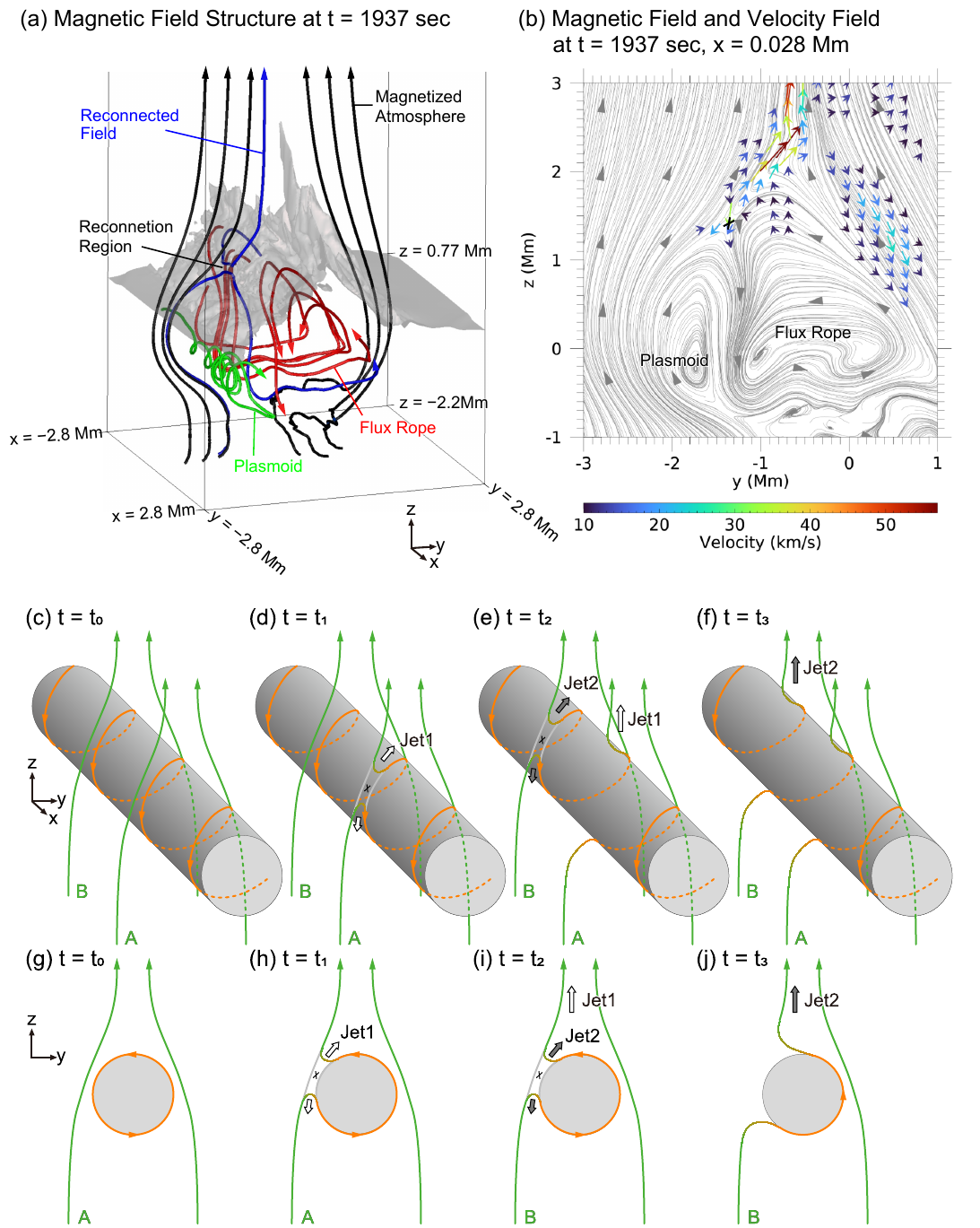}
\caption{(a) The magnetic field structure at $t = 1.94 \times 10^3$~s obtained from the simulation data. The iso-surface of density is shown by grey colour.(b) The magnetic field and velocity field are shown at $x=0.028$~Mm and $t = 1.94 \times 10^3$~s. The magnetic field and velocity field, obtained from simulation data, are shown in the panel (b) here (i.e. Magnetic field and velocity field at $t = 1.94 \times 10^3$~s and $x=0.028$~Mm) correspond to the box shown in Fig. 3c.  The schematic diagrams (c to j) show the origin and evolution of the jets, the oblique view (c to f) and axial view (g to j) are provided in the diagrams. The magnetic flux rope is shown by grey colour. The sign 'X' in the schematic diagram shows the reconnection point, the flux rope magnetic field is shown in orange color and the ambient magnetic field is shown in green color. At first, Jet1 is formed and magnetic reconnection proceeds further leading to more jets such as Jet2 and so on. In this manner, many small-scale jets would be launched from the entire body of the magnetic flux rope.}
\end{figure}
\end{document}